\documentclass[11pt]{amsart}
\usepackage{geometry}                
\usepackage{graphicx}
\usepackage{amssymb}
\usepackage{epstopdf}
\usepackage{breqn}
\title{A Degenerate Bogdanov-Takens Normal Form for FLRW Cosmologies}
\author{Ikjyot Singh Kohli}
\author{Michael C. Haslam}
\address{Department of Mathematics and Statistics, York University, Toronto, Ontario}
\address{Department of Mathematics and Statistics, York University, Toronto, Ontario}
\email{isk@mathstat.yorku.ca}
\email{mchaslam@mathstat.yorku.ca}
\date{July 4, 2016}                                           

\begin{document}

\begin{abstract}
In this paper, we first show that the Einstein field equations for all perfect-fluid FLRW cosmologies can be written as a planar dynamical system with the equation of state parameter $w$ and cosmological constant $\Lambda$ as parameters. An important equilibrium point of this dynamical system is the origin which represents Minkowski spacetime, when $\Lambda = 0$. It is shown that the Einstein field equations in a neigbourhood of this point are equivalent to a degenerate Bogdanov-Takens normal form. This normal form admits a set of equilibrium points that describes a set of solutions to the Einstein field equations that have a constant rate of expansion, negative spatial curvature, zero cosmological constant and dust.
\end{abstract}

\maketitle

\section{Introduction}
One of the most important properties of mathematical cosmology are the symmetries of the solutions that arise from solving Killing's equations in conjunction with The Einstein field equations. In particular, for cosmological models that are homogeneous on three-dimensional orbits, the Einstein field equations reduce to ordinary differential equations. Such cosmological models include The Friedmann-Lema\^{i}tre-Robertson-Walker (FLRW) models, spatially homogeneous LRS models, and the Bianchi models \cite{ellis}. Because of the emergence of these coupled ordinary differential equations, several connections over the past number of years between dynamical systems theory and relativistic cosmology have been made. The interested reader is asked to refer to \cite{ellis, elliscosmo} and references therein for a very detailed summary of the literature that clearly shows this deep relation.

An important area of study in dynamical systems theory that has been of some significance in mathematical cosmology is bifurcation theory. Bifurcations naturally arise in the aforementioned cosmological models. This is because the vast majority of these models have at minimum a free parameter representing some physical property of the matter content that the model describes. As such, variations in these parameters lead to changes in stability of the equilibrium points of the corresponding dynamical system. Since equilibrium points in these cases are also solutions to Einstein's field equations, knowing the stability properties, in particular, how these stability properties change with respect to physical parameters is very important from a physical perspective. 

In this paper, we are primarily concerned with the Bogdanov-Takens bifurcation, and analyzing whether such a bifurcation occurs in the present context. To the best of the authors' knowledge, Bogdanov-Takens bifurcations have only been demonstrated to occur in the cosmological context in \cite{Kim:2013xu}, where the authors studied in detail, the chaotic dynamics of the Bianchi IX universe in Gauss-Bonnet gravity. 

In what follows, we shall investigate the existence of a Bogdanov-Takens bifurcation in perfect-fluid FLRW cosmologies. Such an investigation has not been carried out in the literature thus far. Note that, throughout, we use units where $c=8\pi G = 1$.

\section{The Dynamical Equations}
A cosmological model must be specified by a pseudo-Riemannian manifold, $\mathcal{M}$, a metric tensor, $g$, and a four-velocity field, $u$ that describes the matter flow in the model. The matter in the cosmological model is related to the spacetime geometry via the Einstein field equations:
\begin{equation}
R_{ab}  - \frac{1}{2}Rg_{ab} + \Lambda g_{ab} = T_{ab}.
\end{equation}
Following \cite{ellis, hervik}, one can then further decompose the covariant derivative of the four-velocity field as
\begin{equation}
u_{a;b} = \frac{1}{3}\theta h_{ab} + \sigma_{ab} + \omega_{ab} - \dot{u}_{a}u_{b},
\end{equation}
where $\theta$ is known as the expansion scalar, $h_{ab}$ is the projection tensor, $\sigma_{ab}$ is the kinematic shear tensor, $\omega_{ab}$ is the vorticity tensor, and $\dot{u}_{a}$ is the four-acceleration field. 

In this paper, we are concerned with FLRW spatially homogeneous and isotropic cosmological models, and as such, we have that $\sigma_{ab} = \omega_{ab} = \dot{u}_{a} = 0$. Further, this implies that the energy-momentum tensor must have the form of a perfect fluid:
\begin{equation}
T_{ab} = \mu u_{a} u_{b} + p h_{ab},
\end{equation}
where $\mu$ represents the energy density and $p$ the pressure. We make the further assumption that the matter in this model obeys a barotropic equation of state, so that we can write $p = w \mu$, where $-1 < w \leq 1$ is an equation of state parameter, and describes different types of matter configurations in the cosmological model under consideration. For example, $w = 1/3$ describes a radiation-dominated universe, while $w = 0$ describes a universe with pressure-less dust. 

The Einstein field equations then imply the energy-momentum conservation equation
\begin{equation}
\label{eq:enmom1}
\dot{\mu} + \theta\left[\mu(1+w)\right] = 0,
\end{equation}
Raychaudhuri equation
\begin{equation}
\label{eq:raych1}
\dot{\theta} + \frac{1}{3}\theta^2 + \frac{1}{2}\left[\mu(1 + 3 w)\right] - \Lambda = 0,
\end{equation}
and Friedmann equation
\begin{equation}
\label{eq:friedmann1}
\frac{1}{3}\theta^2 = \mu + \Lambda - \frac{1}{2} \mathcal{R},
\end{equation}
where $\mathcal{R}$ denotes the three-dimensional Ricci scalar (related to the spacetime foliation) and $\Lambda$ denotes the cosmological constant.

In fact, upon differentiating Eq. \eqref{eq:friedmann1} with respect to time, Eqs. \eqref{eq:enmom1}-\eqref{eq:friedmann1} imply the following planar dynamical system:
\begin{eqnarray}
\label{eqsys1}
\dot{\theta} &=& \frac{1}{4} \left(-\mathcal{R} - 3 \mathcal{R} w - 2 \theta^2 - 2 w \theta^2 + 6 \Lambda + 6 w \Lambda\right), \\
\label{eqsys2}
\dot{\mathcal{R}} &=& -\frac{2}{3}\mathcal{R}\theta.
\end{eqnarray}
Together, Eqs. \eqref{eqsys1} and \eqref{eqsys2} fully describe the dynamics of all perfect-fluid FLRW cosmological models. A similar methodology in deriving these equations was used in \cite{goliathellis}. Let us write these equations as a sum of linear and nonlinear and constant terms as follows:
\begin{equation}
\label{eq:sysmform}
\left(
\begin{array}{c}
 \dot{\theta}  \\
 \dot{\mathcal{R}} \\
\end{array}
\right) = -\frac{1}{4}\left(
\begin{array}{cc}
 0 & 3w + 1 \\
 0 & 0 \\
\end{array}
\right) \left(
\begin{array}{c}
 \theta  \\
 \mathcal{R} \\
\end{array}
\right) -\left(
\begin{array}{c}
 \left[\frac{3 w \Lambda }{2}+\frac{3 \Lambda }{2}\right]+\theta ^2 \left[\frac{w}{2}+\frac{1}{2}\right] \\
 \frac{2}{3}  \mathcal{R}\theta \\
\end{array}
\right).
\end{equation}

The equilibrium points of the system \eqref{eq:sysmform} are found to be
\begin{equation}
\label{eq:eqpoints}
\left(\theta^{*}, \mathcal{R}^{*}\right) : \left(0, \frac{6 (\Lambda +\Lambda  w)}{3 w+1}\right), \quad \left(\pm \sqrt{3 \Lambda}, 0\right).
\end{equation}

\section{The Bogdanov-Takens Bifurcation}
Let us now consider the case where $\Lambda = w = 0$. One sees that Eq. \eqref{eq:sysmform} becomes
\begin{equation}
\label{eq:sysmform2}
\left(
\begin{array}{c}
 \dot{\theta}  \\
 \dot{\mathcal{R}} \\
\end{array}
\right) = -\frac{1}{4}\left(
\begin{array}{cc}
 0 & 1 \\
 0 & 0 \\
\end{array}
\right) \left(
\begin{array}{c}
 \theta  \\
 \mathcal{R} \\
\end{array}
\right) - \left(
\begin{array}{c}
 \frac{\theta ^2}{2} \\
 \frac{2 \mathcal{R} \theta }{3} \\
\end{array}
\right).
\end{equation}
One sees immediately that the Jacobian matrix in Eq. \eqref{eq:sysmform2} is precisely the zero Jordan block of order 2, and therefore, implies the existence of a Bogdanov-Takens / double zero-eigenvalue bifurcation \cite{kuznet}.

In fact, from Eq. \eqref{eq:eqpoints}, we see that for $\Lambda = w = 0$, the system \eqref{eq:sysmform} has an equilibrium point given by
\begin{equation}
\left(\theta^{*}, \mathcal{R}^{*}\right) = \left(0,0\right),
\end{equation}
which is precisely the origin of the phase space of the system described by Eq. \eqref{eq:sysmform}.

Following Theorem 8.5 in \cite{kuznet}, we note that any generic planar two-parameter system $\dot{x} = f(x,\alpha)$, having at $\alpha = 0$, an equilibrium $x=0$ with two zero eigenvalues, which is known as the Bogdanov-Takens condition, is locally topologically equivalent in a neigbourhood of $x=0$ to the following normal form:
\begin{eqnarray}
\label{eq:eta1}
\dot{\eta}_{1} &=& \eta_{2}, \\
\label{eq:eta2}
\dot{\eta}_{2} &=& \beta_{1}\eta_{1}^2 + \beta_{2} \eta_{1} \eta_{2}.
\end{eqnarray}
Note that, for convenience in what follows below, we have used the form of the Bogdanov-Takens bifurcation as defined in \cite{Peng2011}. Further, this normal form is nondegenerate if $\beta_{1} \beta_{2} \neq 0$. As long as this nondegeneracy condition is obeyed, then, it is possible to derive a generic Bogdanov-Takens normal form for the associate dynamical system. As we shall see below, Einstein's equations \eqref{eqsys1}-\eqref{eqsys2}, violate this condition, and so it is only possible to define a degenerate Bogdanov-Takens normal form, which in this case, is simply termed the critical Bogdanov-Takens normal form.

We will now show that the system \eqref{eq:sysmform2} can be written in the form of Eqs. \eqref{eq:eta1}-\eqref{eq:eta2} via the following coordinate transformations:
\begin{equation}
\label{eq:transform1}
z_{1} = \theta, \quad z_{2} = -\frac{1}{4}\mathcal{R} - \frac{1}{2}\theta^2.
\end{equation}
Indeed, under this coordinate transformation, one obtains
\begin{eqnarray}
\dot{z}_{1} &=& z_{2} + H.O., \\
\dot{z}_{2} &=& -\frac{5}{3}z_{1}z_{2} + H.O.,
\end{eqnarray}
where $H.O$ indicates higher-order terms, which according to Lemma 8.8 in \cite{kuznet} can be ignored, thus giving
\begin{eqnarray}
\label{eq:transform2}
\dot{z}_{1} &=& z_{2}, \\
\label{eq:transform22}
\dot{z}_{2} &=& -\frac{5}{3}z_{1}z_{2}.
\end{eqnarray}

\section{A Further Analysis of the Normal Form}
\begin{figure}[h]
\centering
\includegraphics[scale=0.90]{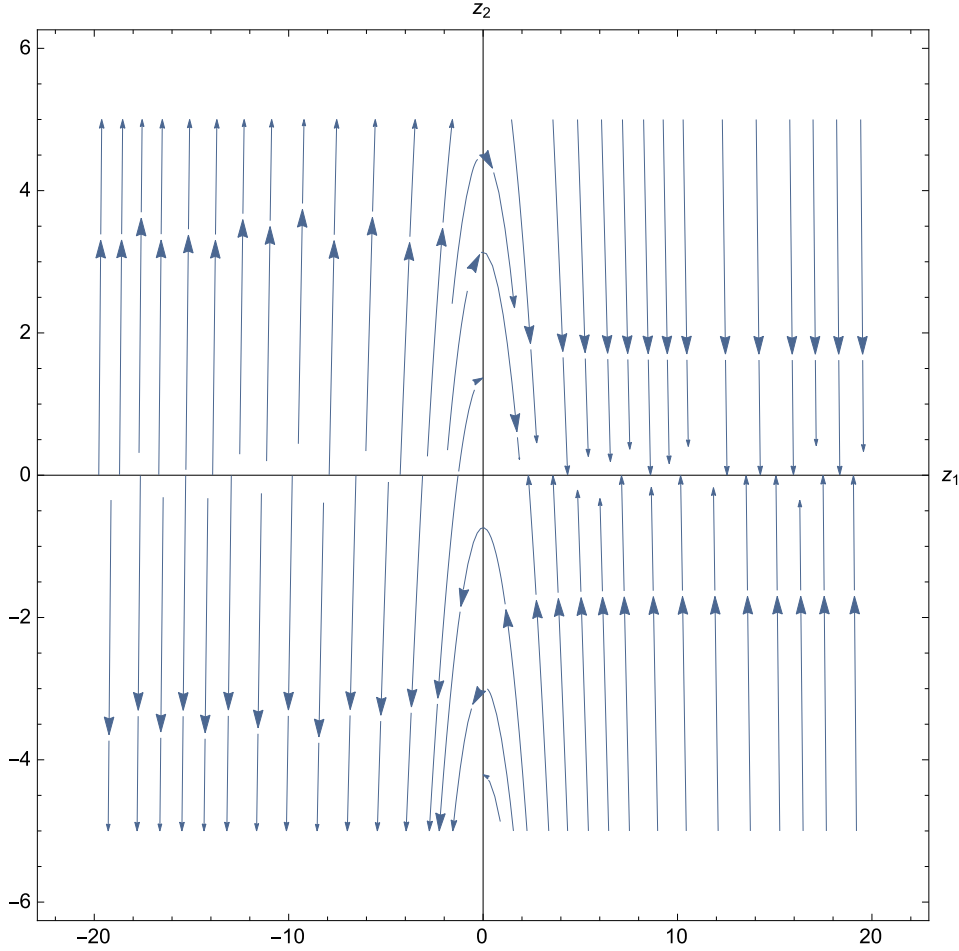}
\caption{A phase portrait of the degenerate normal form for the Bogdanov-Takens bifurcation corresponding to Eqs. \eqref{eq:transform2}-\eqref{eq:transform22}.}
\label{fig:fig1}
\end{figure}

With Eqs. \eqref{eq:transform2} and \eqref{eq:transform22} in hand, we will analyze in this section, the implications of this degenerate normal form further. First, note that the equilibrium points are given by the set:
\begin{equation}
\label{eq:eqpoint1}
z_{1} = c, \quad z_{2} = 0, \quad c \in \mathbb{R}.
\end{equation}
Further, the eigenvalues corresponding to this line of equilibrium points are found to be:
\begin{equation}
\label{eq:eigs1}
\lambda_{1} = 0, \quad  \lambda_{2} = -\frac{5}{3}c.
\end{equation}
Note that, despite the presence of the zero eigenvalue, the point represented by Eq. \eqref{eq:eqpoint1} is normally hyperbolic \cite{ellis}, and one can examine the stability of this point using $\lambda_{2}$ alone. In particular, we have that this point is a stable node if $c > 0$, while it is an unstable node if $c < 0$. When $c=0$, then the point is a Bogdanov-Takens equilibrium point, which is non-hyperbolic, but unstable \cite{kuznet}. In Figure \ref{fig:fig1}, we display a phase portrait of Eqs. \eqref{eq:transform2}-\eqref{eq:transform22} showing this behaviour in detail. 
\newpage

Looking at Eq. \eqref{eq:transform1}, one sees that from a cosmological perspective, we see that the set of equilibrium points defined by Eq. \eqref{eq:eqpoint1} imply that in the original coordinates, we have that
\begin{equation}
\label{eq:params3}
\theta = c, \quad \mathcal{R} = -2c^2.
\end{equation}
That is, the set of equilibrium points defines a set of solutions to the Einstein field equations which describe universes that have a constant rate of expansion, negative spatial curvature, zero cosmological constant and dust, when $c \neq 0$. From our stability analysis of the normal form above, we see that expanding solutions are stable, while contracting solutions are unstable.

For the case $c=0$, Eqs. \eqref{eq:params3} and \eqref{eq:friedmann1}, imply that the solution to Einstein's field equations in this case is trivially Minkowski spacetime.

\section{Conclusions}
In this paper, we showed that the Einstein field equations for all perfect-fluid FLRW cosmologies can be written as a planar dynamical system with the equation of state parameter $w$ and cosmological constant $\Lambda$ as parameters. An important equilibrium point of this dynamical system is the origin, which represents Minkowski spacetime, when $\Lambda = 0$. It was then shown that the Einstein field equations in a neigbourhood of this point were equivalent to a degenerate Bogdanov-Takens normal form. This normal form itself was found to admit a set of equilibrium points that describe a set of solutions to the Einstein field equations that have a constant rate of expansion, negative spatial curvature, zero cosmological constant and dust. Through a stability analysis of these equilibrium points, it was found that expanding solutions are stable, while contracting solutions are unstable.

\section{Acknowledgements}
This research was partially supported by a grant given to MCH from the Natural Sciences and Engineering Research Council of Canada.

\newpage
\bibliographystyle{ieeetr}
\bibliography{sources}

\begin{thebibliography}{1}

\bibitem{ellis}
J.~Wainwright and G.~Ellis, {\em Dynamical Systems in Cosmology}.
\newblock Cambridge University Press, first~ed., 1997.

\bibitem{elliscosmo}
G.~F. Ellis, R.~Maartens, and M.~A. MacCallum, {\em Relativistic Cosmology}.
\newblock Cambridge University Press, first~ed., 2012.

\bibitem{Kim:2013xu}
E.~J. Kim and S.~Kawai, ``{Chaotic dynamics of the Bianchi IX universe in
  Gauss-Bonnet gravity},'' {\em Phys. Rev.}, vol.~D87, no.~8, p.~083517, 2013.

\bibitem{hervik}
{\O}.~Gr{\o}n and S.~Hervik, {\em Einstein's General Theory of Relativity: With
  Modern Applications in Cosmology}.
\newblock Springer, first~ed., 2007.

\bibitem{goliathellis}
M.~Goliath and G.~F.~R. Ellis, ``Homogeneous cosmologies with a cosmological
  constant,'' {\em Phys. Rev. D}, vol.~60, p.~023502, May 1999.

\bibitem{kuznet}
Y.~A. Kuznetsov, {\em Elements of Applied Bifurcation Theory}.
\newblock Springer-Verlag, third~ed., 2004.

\bibitem{Peng2011}
G.~Peng and Y.~Jiang, ``Practical computation of normal forms
  of the bogdanov-takens bifurcation,'' {\em Nonlinear Dynamics}, vol.~66,
  no.~1, pp.~99--132, 2011.

\end{thebibliography}

\end{document}